\documentclass[12pt,preprint]{aastex}

\def\etal{{\it et~al.~}}
\def\nt{{nonthermal~}}
\def\bsax{{\it BeppoSAX~}}

\def\ginga{{\it Ginga~}}

\def\rxte{{\it RXTE~}}

\begin{document}

\newcommand{\lessim}{\ \raise -2.truept\hbox{\rlap{\hbox{$\sim$}}\raise5.truept
    \hbox{$<$}\ }}

\title{Comments to the review " Nonthermal phenomena in clusters of
galaxies" by Y.Rephaeli et al. that will appear on the book:
Clusters of galaxies: beyond the thermal view.}

\author{Roberto Fusco-Femiano$^a$, Mauro Orlandini$^b$
\footnote{$^a$Istituto di Astrofisica Spaziale e Fisica Cosmica
(IASF/Roma), INAF, via del Fosso del Cavaliere, I--00133 Roma,
Italy - roberto.fuscofemiano@iasf-roma.inaf.it; $^b$IASF/Bologna,
INAF, via Gobetti 101, I--40129 Bologna, Italy -
orlandini@iasfbo.inaf.it}}


\affil{}

\par\noindent
\par\noindent
When the review on \nt phenomena in clusters of galaxies by
Rephaeli, Nevalainen, Ohashi \& Bykov 2008 (astro-ph/08010982;
hereafter, RNOB08) appeared on the WEB we sent our comments to
Prof. Y.Rephaeli. But the answer of the Editor, Dr J.Kaastra, of
the book: "Clusters of galaxies: beyond the thermal view", was
that he has checked with Springer, but unfortunately their process
is already too far to make any changes to the paper. For this
reason we have decided to put on the WEB these comments.

\par\noindent
The comments regard: \textbf{a)} the boring controversial between
the analysis of the PDS/\bsax \\ data of the Coma cluster with the
software package XAS by Fusco-Femiano \etal 2004 (hereafter, FF04)
and the Rossetti \& Molendi analysis with a different software
SAXDAS (hereafter, RM04);  \textbf{b)} the  a hard excess in
A2199, A2163 and the Bullet cluster.

\par\noindent
\textbf{Coma cluster:} In 2007 Fusco-Femiano, Landi \& Orlandini
(hereafter, FF07) have re-analyzed the PDS data using the same
software of RM04 showing that it is possible to obtain the same
results of FF04 explaining of course the reasons of the
discrepancy between FF04 and RM04. Rossetti \& Molendi replied to
our paper (FF07) with an electronic preprint only (RM07) and we
were obliged to a new reply (FF07R).

\par\noindent
Unfortunately, the authors of the review have not read with the
due attention the papers FF04, FF07 and FF07R (the last is not
reported in the review) and this is a serious mistake for people
that intend to write a review. So, we are obliged to repeat here
briefly some of the things that are contained in the above papers.

\par\noindent
In FF07 and in the reply FF07R we have reported that to explain
the discrepancy between FF04 and RM04 a rigorous selection of the
events is necessary in order to eliminate the presence of any
spikes able to introduce noise that hides the presence of a \nt
excess with respect to the thermal radiation. We have a
significant increase of the c.l. of the excess (from $\sim
2.9\sigma$ to $\sim 4.2\sigma$) when we consider in the SAXDAS
analysis (FF07) the same time windows used in the XAS analysis
(FF04). The authors of the review report instead that the
discrepancy is \textbf{only} due to the different determination of
the background between FF04 and RM04. In FF04 we consider only the
-OFF direction for the presence of a contaminating BL Lac object
in the +OFF direction. Besides, the authors of the review omit to
report that in FF04 and FF07R we have shown that also considering
the standard technique that implies an average of the two
backgrounds determinations, the c.l. of the excess is still at
$3.9\sigma$.

\par\noindent
The authors of the review report that a point raised by RM07 to
defend their re-analysis and the lower detection significance of
the hard excess was also the choice in FF04 of the value of the
temperature ($8.11\pm 0.07$ keV measured by \ginga, David \etal
1993). Following RM04 and RM07 a more appropriate value is
$8.21\pm 0.16$ keV. RBNO08 omit to report in their review that in
FF07R we have computed, to satisfy Rossetti \& Molendi, the excess
assuming a gas temperature of 8.4 keV, the upper limit of the
reported interval in RM04 and RM07. We obtain a c.l. for the
excess of $4.15\sigma$. Moreover, the \ginga value has been
confirmed by \rxte that reports $7.90\pm 0.03$ keV (Rephaeli \&
Gruber 2002) and in the fit of Fig. 1 of the review the \rxte data
give 7.67 keV (!!).

\par\noindent
The authors of the review report that we have never clarified the
point regarding the possible presence of systematic errors raised
by RM04 and RM07. In FF07 and in FF07R we have stressed that the
systematic errors are discussed in detail by Fusco-Femiano, Landi
\& Orlandini 2005 (FF05) for the analysis of the excess in A2256.
Our surprise for this statement present in the review is that the
referee of FF05 was Dr J.Nevalainen, one of the authors of the
review. In particular, the referee was in agreement with our
analysis on the \textit{whole} sample of PDS pointings (869, while
RM04 consider only 69 observations) regarding the possible
systematic difference between the OFF fields reported in RM04 and
RM07. Our analysis gives a value of $(5.3\pm 6.3)\times 10^{-3}$,
consistent with no contamination at all. Besides, the same sample
was used to measure the X-ray background (Frontera \etal 2007) and
the results are absolutely consistent with the Integral results
with a PDS flux that is lower of $\sim 10\%$ of the Integral flux.
This a further confirmation of the correctness of the PDS results.

\par\noindent
\textbf{A2199:}for this cluster the authors of the review probably
ignore that Fusco-Femiano \etal (2003) have re-analyzed the MECS
data showing that the \nt excess reported by Kaastra \etal (1999)
is not present. The discrepancy is probably due to the use of a
more evolved software package by Sabrina De Grandi.

\par\noindent
\textbf{A2163:}the authors of the review do not report in Sect.
3.4 "Search for NT emission with \bsax" that the PDS observation
gives only an upper limit to the \nt flux as reported by Feretti
\etal (2001).

\par\noindent
\textbf{A2163 \& Bullet cluster:} Finally, we have expressed to
the authors of the review our invite to avoid to present the
excess in these two clusters reported by \rxte observations as
firm detections considering the large error bars in the spectra.

\par\noindent
Even if the book editor, Dr. J.Kaastra, affirms that the
publishing process is already too far, we would have same doubt to
publish a review that contains uncorrect statements.

\par\noindent
We think that, for the sake of intellectual honesty, at least an
addendum page should be included in the book taking into account
our comments, in order to allow the reader to have a more complete
grasp on one of the book main topics: non-thermal phenomena in
clusters of galaxies.

\end{document}